\documentclass[authoryear,12pt,3p]{jowarticle}

\usepackage[english]{babel}
\usepackage[utf8]{inputenc}
\usepackage{amsmath,amsfonts,amssymb}
\usepackage{graphicx}
\usepackage[colorinlistoftodos]{todonotes}
\usepackage{natbib}
\usepackage{blindtext}
\usepackage{subcaption}
\usepackage{caption}
\usepackage{setspace}
\usepackage[shortlabels]{enumitem}

\makeatletter
\renewcommand\section{\@startsection{section}{1}{\z@}{-3.25ex plus -1ex minus -.2ex}{1.5ex plus .2ex}{\normalsize\bf}}
\renewcommand\subsection{\@startsection{subsection}{2}{\z@}{-3.25ex plus -1ex minus -.2ex}{1.5ex plus .2ex}{\normalsize\bf}}
\renewcommand\subsubsection{\@startsection{subsubsection}{3}{\z@}{-3.25ex plus -1ex minus -.2ex}{1.5ex plus .2ex}{\normalsize\bf}}
\makeatother

\begin{document}
\begin{frontmatter}
\title{Spacetime Models for Cosmology}
\author{James Owen Weatherall}\ead{james.owen.weatherall@uci.edu}
\address{Department of Logic and Philosophy of Science \\ University of California, Irvine}

\date{\today}

\begin{abstract}
I revisit Roberto Torretti's ``Spacetime Models for the World'' [\emph{SHPMP} 31 (2):171-186 (2000)] in the light of more recent work in (philosophy of) cosmology.  I discuss the motivations for FLRW spacetimes as a natural starting point for inquiry, and I suggest contemporary cosmologists can avoid the rationalism that Torretti attributes to Einstein's early work in relativistic cosmology.  I then discuss the senses in which FLRW models are idealized, and I show how those idealizations (and partial de-idealizations) have contributed to our understanding of the universe.
\end{abstract}
\end{frontmatter}

\doublespacing
\section{Introduction} \label{sec:intro}

As with many things, modern philosophy of cosmology emerged as an autonomous subdiscipline of philosophy of science (or physics) in two ways: gradually, and then suddenly.  A notable change occurred at the turn of the 21st century, around the same time that the $\Lambda$CDM model came to be accepted as the ``Standard'' or ``Concordance'' model of cosmology.  By ``modern'' philosophy of cosmology, I have in mind work that moves beyond both very general questions about the possibility of cosmology as a science (e.g., Kant, \citep{McMullin}) and technical issues in the foundations of general relativity and Newtonian gravitation bearing on cosmological models \citep[e.g.][]{EarmanBCWS,MalamentNewtCos,NortonNewtCos}, to more directly engage with the substance and methods of contemporary cosmology.  Among the papers that marked the transition, with some modern elements admixed with distinctly pre-modern ones, was Roberto Torretti's ``Spacetime Models for the World'' \citeyear{Torretti}.  In it, Torretti, a giant of twentieth century philosophy of space and time, considers what it means to apply modern physics, and especially general relativity and quantum mechanics, to the universe as a whole.

Among the most striking -- and strikingly pre-modern -- features of the paper is a broad pessimism about the very coherence of the Standard Model of cosmology.  It is not often that one encounters skepticism about an entire field of science in a mainstream philosophy of physics journal, much less from such an accomplished philosopher of science.  Unsurprisingly, the skepticism is rooted in a deep understanding of general relativity, quantum mechanics, and the history and philosophy of physics.  And yet, as I will argue here, there are good reasons to think the skepticism is misplaced.  These reasons are themselves illuminating, both with respect to the subsequent development of the philosophy of cosmology and the status of the methodological considerations Torretti raises.

I will focus, here, on two arguments Torretti makes in the paper.\footnote{There are other arguments as well, such as that early universe cosmology involves mixing general relativity and quantum mechanics in a way that neglects the conceptual incompatibility of those theories.  But I will not discuss them here.}  The first is that relativistic cosmology suffers from an original sin -- or, to avoid the genetic fallacy, a fundamental and unavoidable tension -- illustrated by Einstein's very first paper on relativistic cosmology \citeyear{Einstein}.  As is well known, Einstein attempts to develop a cosmological model within general relativity, and then, arguing on broadly \emph{a priori} grounds that the universe must be static, introduces the cosmological constant to his eponymous equation to secure a model that meets his desiderata.\footnote{By \emph{a priori}, here and throughout this manuscript, I truly mean \emph{a priori}: that is, rationalistic arguments given independent of experience and not subject to empirical revision.  This is the kind of reasoning I take Torretti to be concerned with, and troubled by.  Granted, it may be uncharitable to say that Einstein \emph{only} gave \emph{a priori} arguments for his model, as he does point to some weak observational considerations.  But when De Sitter objects to these their later correspondence, Einstein retreats to more rationalistic arguments.  So perhaps it is not so unfair after all.  For much more detail on Einstein's reasoning here, and his correspondence with De Sitter, see \citet{Belot}.  I am grateful to James Read and Chris Smeenk for help on these points.} Many authors have focused on Einstein's subsequent self-flagellation for failing to predict the expansion of the universe as measured by Hubble just a decade later.  But Torretti focuses on a different problem, which is the role that \emph{a priori} conditions apparently play in any cosmological model-building.  As evidence, he cites the fact that although Einstein's static model has been rejected, contemporary cosmology is based on models -- namely, Friedman-Lema\^itre-Robertson-Walker (FLRW) family of models -- that are arguably grounded in similar principles.

Worse, though one could perhaps argue that the models we use in cosmology are merely highly idealized, rather than \emph{a priori}, Torretti suggests that they are resistant to de-idealization.  (This is the second argument I will consider in what follows.)  The problem, as Torretti identifies it, is that the methods of mathematical physics developed by the likes of Galileo, Newton, and their followers, work well for piecemeal application to small systems -- including say, the solar system -- where we can get control over the ways in which a model merely approximates the behavior of the system and introduce modifications and complications to improve the approximation by making the models more realistic.  This is more challenging for the universe as a whole, where, Torretti suggests, de-idealization would require incorporating information about the full complexity of the totality of things into a model of general relativity.  At very least, he argues, he does not detect within contemporary cosmology examples of de-idealization of the sort he would expect if de-idealization were possible.

In the quarter century since Torretti's paper appeared, there has been considerable progress in theoretical and observational cosmology.  There has also be much more development of philosophy of cosmology.  Here I will revisit Torretti's two arguments in light of that development.  First, I will argue, following \citet{Smeenk+Weatherall}, that there are methodological and theoretical reasons, bolstered by strong observational arguments, to use FLRW models in cosmology.  Considerations that Torretti worries are \emph{a priori} are both defeasible and firmly grounded in \emph{a posteriori} facts.  Then I will discuss three senses in which FLRW models may be viewed as idealized.  Torretti appears to have one of these in mind; highlighting the other two shows more clearly how FLRW models can provide new knowledge of the structure and evolution of the universe, much like idealizations do elsewhere in physics.

\section{Original Sin}

As Torretti writes, ``the first spacetime model of the world ever to be proposed by a physicist'' was Einstein's static universe model, presented in his famous ``Cosmological Considerations'' \citeyear{Einstein}.  This is a model of general relativity set on the manifold $\mathbb{R}\times \mathbb{S}^3$, with matter content consisting of a dust field whose 4-velocity determines a congruence of timelike geodesics with vanishing geodesic deviation.  More informally: matter is represented as uniformly distributed through space and non-interacting, moving at a uniform velocity with no relative acceleration, i.e., so that each little speck of matter remains at a constant distance over time from each of its neighbors.  The model is highly symmetric: not only is it uniform (homogeneous and isotropic) at each moment of time, as determined relative to the cosmic dust field, it is also symmetric over time, in the sense that each moment is the same as each other moment.  There is no change in this model, no expansion or contraction.  This is the sense in which it is a \emph{static} model.

There is no question that Einstein's model was a landmark in the history of physical cosmology.  But for Torretti's purposes, the important issue concerns the paper's motivations, rather than its impact.
\begin{quote}\singlespacing [W]e would naturally wish to know what motivated Einstein to come forward with it.  Was this sudden concern for the overall structure of spacetime induced by some new experimental results?  And, if that was not the case, what on earth prompted Einstein's unexpected lunge for totality? ... [I]t does not appear that Einstein's move towards cosmology was driven by the pressure of facts. Nor did Einstein ever suggest it was. \citep[p. 175-6]{Torretti}\end{quote}
Instead, Torretti observes, the motivation was a clash of principles.  The previous year, Schwarzschild had provided the first non-trivial exact solution to Einstein's equation---what we now know eponymously as Schwarzschild spacetime.  He did so by assuming certain symmetry conditions and by imposing a boundary condition of asymptotic flatness.  Einstein had no objections to the symmetry assumptions.  But he rejected asymptotic flatness, because he felt it conflicted with Mach's principle: as Einsteins states it, ``The [metric] field is exhaustively determined by the masses of bodies'' \citep[quoted in][p. 177]{Torretti}.

The goal in introducing a cosmology, then, was to show how his theory was compatible with a model of the world in which there is no such thing as space infinitely far from gravitating matter, where spacetime reverts to a ``default'' flat geometry, independent of matter.  He did so by finding a solution in which space is compact, thereby obviating the need for boundary conditions at infinity.  Under those circumstance, he claimed, the metric everywhere could be viewed as determined by matter, which Einstein modeled as homogeneously and isotropically distributed throughout (finite, compact) space.

As Torretti observes, Einstein's model is fully determined by three assumptions: that ``space'', properly construed, is compact; that matter, and thus the metric, is spatially homogeneous and isotropic; and that the metric is static, in the sense that there is a timelike Killing field (orthogonal to slices of ``space'').  None of these assumptions was even motivated by empirical considerations, much less forced on Einstein by observation.   Instead, Torretti suggests, the first assumption is wholly motivated by an ideological commitment to a particular form of Machianism and the second is supported by some combination of a desire for mathematical tractability and a ``Copernican'' argument that the universe elsewhere is generally the same as how it appears to us.  Similar arguments might be given for the third condition---though, as Einstein would learn a decade after he published his paper, it is incompatible with observations about the motions of distant galaxies, which do not remain fixed over time.  To make matters worse, Einstein was so committed to these ``principles'' that he was willing to modify the field equation for the metric in general relativity that he had introduced two years previously, adding the infamous cosmological constant term to find a solution that met all three desiderata.  Without that term, no such solution exists.

Torretti takes two major lessons from Einstein's example.  The first is that cosmological models are motivated, primarily or entirely, by rationalist commitment to \emph{a priori} principles.  This is true for Einstein's model, but he suggests it is equally true for the modern ``Standard Model'' of cosmology, based on FLRW spacetimes.  FLRW models are also fixed by a symmetry principle, namely, that matter, and thus the metric, is spatially homogeneous and isotropic, relative to some preferred foliation of spacetime.  Indeed, one can see the Einstein solution as a special case of FLRW spacetime, where one adds the assumptions of spatial compactness and the additional timelike symmetry.  When FLRW spacetime was first discovered -- perhaps by Einstein himself in 1917, Torretti suggests, given how he describes his own model, but certainly by the early 1920s by Alexander Friedmann -- the weaker symmetry principles had no better empirical justification than Einstein's other assumptions.  Moving from the static model to a more general FLRW spacetime, from this perspective, is simply a retreat to a more general case, still motivated by mathematical tractability and a Copernican principle, in the face of falsifying evidence.

This leads to the second moral that Torretti draws, which is that the principles on which cosmological models are based can fail in the face of new empirical information---even though it is hard to see how those same observations could ever justify such principles, much less pin down a unique model.  At best, one can find models that are compatible, to some degree of approximation, with known astrophysical results.  He approvingly quotes H. P. Robertson, who claims in a 1933 review article on relativistic cosmology, ``We hope to show that, under the guidance of a few seemingly natural assumptions and extrapolations, we can arrive at an intrinsically reasonable system of relativistic cosmology which is not in serious conflict with modern astrophysics'' \cite[quoted in][p. 181]{Torretti}---approvingly, not because he endorses Robertson's goals, but rather because he finds the description of what is, and could be, achieved, to be accurate.  But ``intrinsically reasonable'' and ``not in serious conflict'' with observations are weak grounds for accepting a scientific theory.  And as Torretti observes, when Robertson wrote his review, there were a half dozen cosmological models available that could be described as meeting the same standards.  There was nothing to decide between them aside from preferences about principles.

The picture that emerges is one on which cosmological theorizing is driven by commitments to \emph{a priori} principles about how the entire world must be, only loosely constrained by observation or standard scientific methodology.  But is that just?  Perhaps it was in the 1930s---though I think a more sympathetic analysis could be given.  But it is certainly not a just description of contemporary cosmology. I will presently argue that even though Torretti is correct that principles play an important role in contemporary cosmology, the justification for the current Standard Model, even while based on FLRW geometry, is far stronger than what he describes.

\section{Justifying cosmological principles}

Torretti correctly emphasizes the role of what he calls ``philosophical desiderata'' -- notably, ``the `Copernican' postulate that we do not hold a privileged position in the universe'' (p. 178) -- in both Einstein's ``Cosmological Considerations'' and in subsequent arguments for adopting an FLRW model as the starting point for the Standard Model of cosmology.  He also correctly argues that these desiderata were not forced on cosmologists in the early twentieth century, and certainly do not hold exactly.  But he does not say much about just what these desiderata are, what motivates them, or what positive arguments might be given in their support.  In what follows, I will attempt to do just that, following arguments that are developed in much more detail in \citep{Smeenk+Weatherall}.

The first thing to observe is that contemporary physical cosmology, as understood by virtually all physicists working on the subject, aims to develop a model of the universe at very large scales, drawing on both our knowledge of (relatively) local physics and astrophysical observations at many scales.  Physicists often talk of ``cosmological theory'', but the goal is not to develop an autonomous theory of cosmology (or of possible universes), so much as to apply other physical theories and known facts from other domains, including gravitational physics, particle physics, thermodynamics, and plasma physics to the early universe and its development at large scales.\footnote{Torretti himself emphasizes the tensions between these domains of physical theorizing, and especially between quantum theory and general relativity.  That tension is real, and it manifests in real physical puzzles about how early universe perturbations are seeded.  But I am much less convinced that it creates the fundamental barriers to physical cosmology that Torretti detects.  In any case, it is beyond the scope of this paper.}  In this, cosmology should be viewed as what Siska de Baerdemaeker calls an ``integrative science'', because it aims to integrate knowledge from many domains to understand the universe at distant epochs and scales \citep{Siska}.

It is true that the result can be viewed as a model of the ``totality'' of things, in something like the sense that Torretti worries is outside the ken of science.  But that perspective needs to be taken with a grain of salt, in two ways.  The first is that it is not necessary for the ordinary activities of theoretical or observational cosmologists that they think of the Standard Model in this way, as opposed to as a model that applies to the universe at the particular (very large) scales that they are able to observe and aim to study; and the second is that, like any modeling activity in theoretical physics, there are aspects of the Standard Model of cosmology that are firmly established, others that are more conjectural, and others still that are implications of the model that may well be viewed as artifacts not to be taken seriously.  In particular, the implications of the model for scales far beyond the observable universe, including the time shortly after the big bang that is observationally inaccessible with current methods, can be and often are kept at arm's length.  One can both ask what the Standard Model implies about the totality while also recognizing that, as a model of totality, it is much less epistemically secure than it is as a model of, say, structure formation after recombination.  Cosmologists certainly understand that.

With these aims of cosmology in mind, consider the principle that Torretti explicitly identifies, viz., the Copernican Principle (CP).\footnote{He also mentioned the symmetry condition that the universe is, on very large scales, spatially homogeneous and isotropic, which is sometimes called the Cosmological Principle.  We return to this presently.}  Torretti offers a standard gloss, which is that the CP says we do not hold a privileged place in the universe.  But what, exactly, does that mean?  What would it mean to deny the CP---and what, if anything, follows from accepting it?

\citet{Smeenk+Weatherall} argue that there are several different ideas that the CP is used to invoke.  One version, perhaps the plain meaning of how the principle is often stated, has a distinctly metaphysical character: one imagines the universe may have a privileged location (e.g., a geometric center or a position of divine favor), but one denies that the Earth occupies that location.  The picture one should have in mind is the shift from geocentrism to heliocentrism: in an older cosmology, the Earth was central (and stationary), and the heavens revolved around it; on the Copernican model, the sun is central and it is the Earth that moves.  Of course, contemporary cosmology is Copernican in this sense, insofar as it is not geocentric (or heliocentric).  But beyond that, this version of the principle is both obscure and apparently very weak.  It is not clear what, if anything, would follow from the Earth being located in a privileged place---or what follows from it not being so-located.\footnote{This version of Copernicanism is sometimes invoked in contemporary cosmology, for instance in arguments that it is implausible to explain the apparent accelerated expansion of the universe by assuming the Earth is located at or near the center of a large void.  But ruling out highly contrived or symmetrical examples of this sort does not translate to positive proposals for a cosmology; moreover, the void case is also ruled out by other versions of the principle described below.}

This version of the CP is often alluded to and criticized.  But it is not often used in any positive way in cosmological theory.  Instead, there are several other principles that apparently play a more substantive positive role.  For instance, \citet{Smeenk+Weatherall} identify the following:
\begin{itemize}
\item Aristotle's Principle: Locally discoverable laws of physics apply elsewhere in the universe, i.e., in different locations and epochs and, to whatever extent is compatible with known physics, on other length, energy, and time scales.
\item Kant's Principle: Observers on Earth occupy a particular perspective on the universe, reflecting the position and state of motion of the Earth, the solar system, and the galaxy relative to other bodies in the universe.  What we observe from our own position may differ from what observers would view from other positions.
\item Law of large volumes: The distribution of physical properties and structure is determined locally by effectively stochastic processes with uniform statistical properties.  While there may be local variation, on sufficiently large scales the distribution and configuration of matter should reflect statistical regularities.
\item No conspiracies: The universe is not finely tuned so as to thwart our attempts at building a physical cosmology.
\end{itemize}
Each of these is in some sense ``Copernican'' in character, and might be seen as a way of making precise the idea that our place in the world is not special.  But they capture different aspects of the idea---and they all emphasize a different sense of ``not special'' than the metaphysical gloss given above.

Broadly, the sense of ``not special'' reflected in these principles is that our position in the world is both \emph{representative} and, in one sense or another, similar to other places. Of course, that similarity need not mean that all places are the same.  As we see in Kant's Principle, it can amount to the observation that \emph{every} position is special, in the sense that the universe viewed from any perspective will be different from other perspectives, but that all of them are the same in the sense that they can be collected in a coherent model that permits imagined perspective switching \citep{Roush}.  Indeed, this version of the CP is arguably closest to the most significant contribution that Copernicus himself made, which was to show that although the Sun appears to move around the Earth, that is precisely what one should \emph{expect} it to look like if in fact the Earth went around the Sun, given certain other facts about physics.  So recognizing that we are viewing the world from our perch on a planet undergoing perfectly mundane motions like any other planet drives home an important consequences of our lack of specialness.

These versions of Copernicanism seem importantly different from the metaphysical one, for two reasons.  The first reason is that they are \emph{methodological} principles, in the sense that they stake out rules of inference and ways of applying other known physics to cosmology.  For this reason, they are also much more fertile than the metaphysical principle.  They suggest directions of research and ways to infer from observations to models of phenomena.  For instance, when we observe that distant stars have similar spectral profiles to the Sun, we infer that they are subject to similar thermonuclear processes.  Likewise, when we observe that distant galaxies rotate at certain angular velocities, or galaxies in clusters have some average relative velocity, we infer that they are gravitationally bound---and thus, that the system must have a certain mass, possibly larger than the mass we measure in other ways.  And so on.  These are all applications of Aristotle's Principle.

Or consider that when we observe the statistical distribution of stars near us, or the distribution of galaxies in the universe, or the temperature of the cosmic microwave background, we will detect characteristic asymmetries due to the fact that the Earth is located towards the outer region of the Milky Way and that it is itself moving, around the Sun and, as part of the solar system, around the center of the galaxy.  This is an application of Kant's Principle, just as Copernicus took into account the motion of the Earth around the Sun when inferring from the observed motions of the planets in the sky to their positions and orbital parameters around the sun.  A suitable application of Kant's Principle also allows us to subtract away our own local motions in drawing inferences about the large-scale statistical distribution of matter in our local region of the universe---which supports an application of the law of large volumes to infer what the even-larger scale statistics of the universe is likely to be.  And finally, the Law of Large Volumes provides a compelling reason to suppose that at sufficiently large scales, the universe should be homogeneous and isotropic, since to the extent that the principle holds, at sufficiently large scales the distribution of matter should reflect the results of sampling from a very large number of independent and identically distributed random variables characterizing the local stochastic processes. For sufficiently well-behaved distributions we would thus expect the statistical properties of different regions of the universe to match one another.

The second difference between these versions of Copernicanism and the metaphysical principle with which we began is that they are underwritten by the aims of physical cosmology.  If the goal of cosmology is build a model of the physical universe on scales much larger than the galaxy, or even our local galaxy cluster, it is very hard to see how we could ever proceed except by supposing that our local universe is representative---representative with respect to the local laws of physics, the local statistics, and the kinds of things one can learn as a situated observer.  Of course, the fact that we apparently need to assume these principles does not make them true.  Perhaps the project of physical cosmology is destined to fail because these principles do not hold.  But then again, if it does fail for that reason, then the fact that we have such a successful history of doing local physics -- local to Earth, to our solar system, and to our galaxy -- immediately raises the question of at what scales these principles begin to fail.  And the way we would probe that would be by assuming the principles, attempting to model the observed phemomena, and seeing what goes wrong.  In other words, it may well be that the local laws of physics are parochial, and that different laws govern the early universe, or distant regions---but if so, it may well also be true that we can come to know that the laws of physics have changed over time, or across places, in the course of doing cosmology.

The result is that a practice of positing models with certain symmetry properties -- in particular, models that are spatially homogeneous and isotropic -- and applying the gravitational dynamics that appear to be well-confirmed at smaller scales is simply to pursue the aims of cosmology, under the assumptions that are apparently necessary for those aims to make sense.  When we do that, we find that the Einstein static model is a possibility, but that it is dynamically unstable, which leads us to the more general family of FLRW models.  From there, we apply Aristotle's Principle and Kant's Principle to try to interpret our astrophysical observations, including the motions of distant galaxies and clusters and the characteristics of the Cosmic Microwave Background.  As we do so, we attempt to refine the model with which we began, to add texture and detail, and to correct errors in our initial modeling---errors such as the assumption that the only types of matter in the universe are those that we have already discovered on Earth.  More importantly, the entire process is defeasible, in the sense that failures to accommodate observations in accordance with the principles laid out can lead cosmologists to revise their commitments.

In other words, far from assuming \emph{a priori} principles to construct a model of the totality of the world, in contradistinction to the methods of mathematical physics that Torretti identifies with the likes of Newton and Galileo, cosmologists proceed by attempting to apply those same methods to larger and larger scales, extrapolating knowledge of local physics to better understand the limits of that knowledge.  The success of that practice is ultimately what justifies both the extrapolation and the principles adopted.

\section{Idealization and De-idealization in Cosmology}

As I have just argued, Torretti is right that principle-based reasoning motivates our cosmological model-building, at least in the first instance. But I have also argued that a clear understanding of what the principles are and how they are used shows that they have a different character from what Torretti suggested.  But this is only to respond to the first of the worries he raises, concerning the practice of introducing models of the totality of the universe on \emph{a priori} grounds.  He also raises a second worry, concerning the role of idealization in cosmology.  Here he argues that cosmologists seem content to work with highly symmetric models on principle-based grounds, even though the most casual observations reveal that the actual universe is highly asymmetric, and least on small scales, as anyone who has looked at the Sun and then away again can see.  He goes on to wonder why cosmologists do not attempt to de-idealize their models---e.g., by attempting to paste together models of smaller scale structures like galaxies or clusters.\footnote{In fact, some cosmologists have attempted just this---including Einstein himself \citep{Einstein+Straus,SwissCheese}.} His conclusion is that this kind of de-idealization is incompatible with cosmology in some way, either because it is misaligned with cosmologists' totalizing goals or because the problem is simply too difficult to carry out in full (i.e., by including all small-scale details) to be fruitful.

It is certainly true that cosmological models are highly idealized---including the Standard Model of Cosmology.  Torretti is also correct that it would be a fruitless exercise to try to build an exact solution to Einstein's equation that incorporated even galactic-scale physics, much less the fine-grained details of stars, planets, and smaller, and that cosmologists do not attempt this.  But that does not mean that cosmologists do not de-idealize---nor that they do not use their idealized models in the same ways as other physicists to gain new knowledge of the world.  To see this, it is helpful to distinguish three types of idealization.\footnote{\citet{Weisberg} also distinguishes three types of idealization, but the distinction here does not track his---and only Galilean idealization appears on both lists.}

First, we have Galilean idealization, which has been widely discussed in the literature \citep[e.g.][]{McMullin,Weisberg}.  Galilean idealization consists of taking a complex system that one does not have the theoretical, mathematical, or computational resources to model in full detail, and representing it as a simpler model.  The principal goal of Galilean idealization is tractability: one simplifies a problem by ignoring known complications, and one argues that the results of analyzing the simplified model will approximate the more complex system one initially wished to study, at least for some purposes.  The standard example to think about is how high school physics textbooks treat wind resistance.  When we drop a mass from some fixed height on earth, it will accelerate due to gravity---but that acceleration will be counteracted, in part, by air pressure.  Since this complicates the problem, and is a relatively small effect for many bodies, we simply ignore it in predicting how the body will fall.  Often, Galilean idealization is useful because it captures important qualitative features of a system, as described by some theory, which allows one to distinguish that theory from others that make different qualitative predictions.  For instance, as Galileo himself emphasized, bodies of different masses will fall from a fixed height in approximately the same time, at least as long as the mass and geometry of the body are such that wind resistance is weak.  This distinguished his theory of motion from the earlier Aristotelian theory.

I wish to distinguish Galilean idealization from two other kinds of idealization.  One is \emph{Newtonian idealization}, as discussed by \citet{SmithCtL, smith2024physical}.  Newtonian idealization, like Galilean idealization, often involves analyzing systems that are simpler than the target.  But the goal of the practice is different.  Newtonian idealization involves attempting to model what one knows about a system, or at least, what features of a system one thinks will be most important for describing its behavior, with the goal of identifying empirical discrepancies. These empirical discrepancies, in turn, can be a source of new knowledge of the target system, because they reveal gaps in one's understanding. The idea is that one gives an exact, or nearly exact, description of what a system would be like were it to have certain perhaps counter-to-fact features, and then one uses that to measure deviations from the modeling assumptions.  Though Smith offers many examples from the history of celestial mechanics, the most famous is the discovery of the planet Neptune, which was predicted on the basis of a deviation in the expected orbital parameters of Uranus.  Newtonian idealization is often most useful in the context of theory-mediated measurement, where one uses a detailed model that incorporates other knowledge about a system to precisely predict how observations will depend on some unknown parameter, and then infer the parameter's value from the results of the observation.

There are two important differences to highlight between Galilean and Newtonian idealization.  First, Newtonian idealization often involves sophisticated methods such as perturbation theory and power series analyses, since even the relatively simple models one wishes to analyze may not have known exact solutions.  (This is often true in cosmology.)  In this sense, a Newtonian idealization may look like a Galilean \emph{de-idealization}, because one begins with a tractable model but then adds further detail in an attempt to capture, with great precision, how the world would be under some specified physical situation---say, a solar system in which there are only seven planets orbiting the sun, with the masses and orbital parameters of our own solar system.  The second difference is that while both Galilean and Newtonian idealization involve neglecting some details, in Galilean idealization one generally can specify what one is neglecting -- wind resistance, for instance -- whereas in Newtonian idealization, the goal is to use the calculation to figure out unknown factors that one has not accounted for.

Both of these types of idealization are to be contrasted with a third type of idealization, which I will call \emph{Platonic idealization}.  Platonic idealization is not widely discussed in philosophy of physics because it is not often employed by physicists, though something like it does appear in the economics literature.\footnote{I have in mind here models of markets as frictionless exchanges, where one conceives of the model market as the true target of analysis, with properties fixed by definition, and real-world markets as approximations to the ideal, rather than the other way around.  See \citet{WeatherallPOWS} for  discussion.}  The idea of Platonic idealization is to identify characteristic, plausibly \emph{a priori} (or at least, analytic) features of an imagined or theoretical target system, and then build a model with those features.  The actual world may not exactly match the model, but that is thought of as a failure on the part of the world, in the same way that the world of becoming is but a shadow of the world of forms.  The goal of the practice is an exercise in conceptual analysis, rather than an attempt to extract further information about the world or make concrete numerical predictions.

When Torretti argues that the modeling practices of cosmologists are problematic, what he seems to have in mind is the claim that cosmologists are engaged in Platonic idealization.  They begin with \emph{a priori} principles and they construct -- on his reconstruction -- a model of the universe that accords with those principles.  If the result does not exactly correspond to the world, so much the worse for the world.  Or at least, he argues, it is not apparently part of the practice to refine the model to better match the world, so much as argue that it does match the world when viewed from a God's eye perspective.  Likewise, when he proposes that a better model for the universe as a whole should at least account for the fact that there are isolated masses in the universe, separated by a void, rather than a uniform cosmic fluid as contemplated by FLRW models, one can interpret him as proposing a Platonic de-idealization: that is, another Platonic idealization guided by assumptions that somewhat more closely match qualitative features of the world, though still not attempting to use the idealization to guide further inquiry.

It is plausible that Platonic idealization is a reasonable description of some early cosmological modeling.\footnote{Though again, see \citep{Einstein+Straus}.  It seems Torretti was not aware of this later work by Einstein when he published his article.  (I am grateful to Chris Smeenk for raising this point.)}  And if it were the case that contemporary cosmologists engaged in that practice, I would share Torretti's worries.  But in fact, cosmologists are better viewed as employing both Galilean and Newtonian idealization, and not Platonic idealization.  This is closely related to the aims of cosmology as discussed above, and, in particular, the idea that cosmologists do not aim to construct a model of totality, so much as use standard mathematical physics methods to learn about the structure of the universe on very large scales.

I will support this claim with three examples, one of Galilean and two of Newtonian idealization in cosmology.  All of the examples begin with an FLRW model as the idealization, but the reasoning in each case is different.  Consider, first, early structure formation in the universe.  In this case, it is clear that the target system involves many gravitationally attracting systems -- the seeds of early galaxies and clusters -- undergoing a complex dynamics.  A detailed analysis of early structure formation is intractable. But cosmologists can nonetheless analyze a simplified version of this system, first by studying FLRW models as a 0th order approximation and then by de-idealizing by incorporating leading order perturbations.  This leads to what is known as first-order (or linear) perturbation theory, which is a useful tool for reasoning about structure in the early universe \citep{Mukhanov+etal}.  I suggest this is best understood as an instance of Galilean (de-)idealization, because one understands that small inhomogeneities will grow, in a non-linear way; but one neglects the higher order effects on the grounds that it is more tractable to study the simpler theory.\footnote{One might respond that this example is not so much Galilean idealization as straightforward \emph{approximation}, in the sense that one can often get quantitative control over the degree to which higher order corrections will be important.  Fair enough---though it would seem this is often true in cases of Galilean idealization, at least eventually, and yet the qualitative features of the simplified system are nevertheless sufficient for the intended application of the idealization. I am grateful to Chris Smeenk for pressing this point.}  This idealization has been important in cosmology, for instance because first-order perturbation theory is sufficient to distinguish the effects of different dark matter models on early structure formation.  Such arguments have been used to rule out dark matter models in which the dark matter is ``hot'', i.e., have large average kinetic energy.

First-order perturbation theory has also figured in instances of Newtonian idealization.  It is clear that first-order perturbation theory provides a simplified representation of the early universe.  Nonetheless, we have strong empirical grounds, from the isotropy of the cosmic microwave background (CMB), to infer that the early universe was very close to spatially homogeneous and isotropic.  This means a description of the universe at that epoch that deviates from an FLRW model only to first order can be highly accurate \citep{Hu+Dodelson}.  Using this theory, cosmologists in the 1990s derived detailed predictions for how a key observable, the spectrum of temperature anisotropies in the CMB, would depend on a variety of cosmological parameters, such as the density of baryonic matter in the universe, the density of dark matter, and the cosmological constant.  Since some of these parameters could be measured independently, and others have different characteristic effects on the data, these predictions were the basis for high-precision theory-mediated measurements of those parameters in the early universe, beginning in 2003 with the release of data from the Wilkinson Microwave Anisotropy Probe (WMAP) \citep{WMAP}.

Finally, we mention one more example of Newtonian idealization and theory mediated measurement, this time from the late universe.  Generally speaking, cosmologists believe that even in the current cosmological epoch, where there is rich and heterogeneous structure on smaller scales, that at very large scales FLRW models provide an accurate representation of the structure of the universe.  This assessment is based on the observed homogeneity of the early universe, as measured from the CMB, along with observed statistical regularities at much later stages.  Assuming FLRW geometry on very large scales allows cosmologists to measure the expansion history of the universe, and particularly the acceleration parameter, by measuring the relative velocities of bodies at different distances from the earth.  Using this technique, in 1998 two groups announced the theory-mediated measurement of an unexpected acceleration in the expansion of the universe, corresponding to a small but positive value of the cosmological constant \citep{Riess+etal,Perlmutter+etal}.  This value could be compared to the WMAP measurements to provide an even more precise measurement.

\section{Conclusion}

Torretti is correct to observe that cosmologists use highly idealized models, built, in part, using principle-based reasoning.  But his pessimism about ``the standard cosmological model and its variants'' (p. 171) is misplaced.  A more careful investigation of the practice of cosmology shows that the principles at work in modern cosmological theorizing do not have the rationalistic, \emph{a priori} character that Torretti suggests; instead, they are best seen as methodological principles adopted to extrapolate from our knowledge of local physics to the structure of the universe on much larger scales.  These inferences can be tested observationally, and have been shown to be well-supported by empirical evidence.  More importantly, cosmological models have been used to make highly precise measurements of cosmological parameters---measurements that require cosmologists to deploy detailed theoretical and observational knowledge of the universe.

\section*{Acknowledgments}
I am grateful to Pablo Acu\~na for organizing the session at CLMPST 2023 where this talk was presented.  Thanks, too, to James Read for comments on a previous draft; to Chris Smeenk for both helpful feedback on this manuscript and for many conversations related to this material; and to Roberto Torretti for his inspiration, generosity, and mentorship.

\bibliographystyle{elsarticle-harv}
\bibliography{cosmology}

\end{document}